\documentclass{cernyrep}
\begin{document}
\title{Giant Liquid Argon Observatory for Proton Decay, Neutrino Astrophysics and CP-violation in the Lepton Sector 
(GLACIER)\footnote{Contribution to the Workshop ``European Strategy for Future Neutrino Physics'', CERN, Oct. 2009, to appear in the Proceedings.}}
\author{A.~Badertscher, A.~Curioni, U.~Degunda, L.~Epprecht, S.~Horikawa, L.~Knecht, C.~Lazzaro, D.~Lussi, A.~Marchionni,
	G.~Natterer, P.~Otiougova\thanks{Now at University of Z\"urich, Physik-Institut, CH--8057 Z\"urich, Switzerland.},
	F.~Resnati, A.~Rubbia\thanks{Corresponding author: andre.rubbia@cern.ch}, C.~Strabel, J.~Ulbricht,~ and T.~Viant}
\institute{ETH Zurich, 101 Raemistrasse, CH-8092 Zurich, Switzerland}
\maketitle

\begin{abstract}
GLACIER (Giant Liquid Argon Charge Imaging ExpeRiment) is a large underground observatory for proton decay  
search, neutrino astrophysics and CP-violation studies in the lepton sector. Possible underground sites are studied
within the FP7 LAGUNA project (Europe) and along the JPARC neutrino beam in collaboration with KEK (Japan).
The concept is scalable to very large masses. 
\end{abstract}

\section{Introduction}

Looking at the future of neutrino long-baseline experiments, T2K \cite{T2K} and NOvA \cite{NOvA} can be 
considered as "Phase I" experiments that will improve the sensitivity to the small mixing angle $\theta _{13}$ 
by one order of magnitude over the existing CHOOZ limit (sin$^2 2\theta_{13}$>0.1) \cite{CHOOZ}, but will
have limited sensitivity to the effects of CP violation in the neutrino sector.
The next generation of "Phase II" experiments is expected to have significant discovery potential for CP violation, 
capability to establish the mass hierarchy and - in case $\theta _{13}$ is small enough to elude "Phase I" experiment - 
extend the $\theta _{13}$ discovery potential. 
In order to do so, a "Phase II" experiment must be designed to have very large statistics, excellent background 
rejection and very good energy resolution, to measure very precisely the oscillation pattern as a function of neutrino
energy. The detector must therefore be very massive (100,000 to 1,000,000 tons of active volume, depending on 
specific detector technology), and must have the capability to accurately reconstruct neutrino interactions for energies
of about 1 GeV. 
A liquid argon time projection chamber (LAr TPC: see for example \cite{T600} and references therein) provides high 
efficiency for $\nu _e$ charged current interactions ("signal" events), with high rejection power against $\nu _{\mu}$ 
neutral and charged currents backgrounds in the GeV and multi-GeV region .
In particular, excellent e / $\pi ^0$ separation comes from fine sampling (few \% of a radiation 
length), and a transverse sampling finer than the typical spatial separation of the 2 $\gamma$'s from 
the  $\pi ^0$ decay. The identification capability for electrons, $\mu$, $\pi$, K, and protons is also excellent down to 
energies as low as few tens of MeVs.
Embedded in a magnetic field, a LAr TPC gives the possibility to measure both wrong sign muons and wrong 
sign electrons samples in a neutrino factory beam \cite{Rubbia:2001pk, Ereditato:2005yx}.
Unlike Water Cherenkov detectors, detection and reconstruction efficiencies do not depend on the volume of
the detector, therefore a direct comparison between the near and far detector is possible (apart from flux extrapolation).
A very large LAr TPC as the one needed in a "Phase II" long baseline neutrino experiment would also have
unprecedented sensitivity to proton decay \cite{Bueno:2007um} and neutrinos from astrophysical and terrestrial 
sources (solar and atmospheric neutrinos, neutrinos from stellar collapse \cite{GilBotella:2004bv, Cocco:2004ac},
or neutrinos from Dark Matter annihilation).

GLACIER (Giant Liquid Argon Charge Imaging ExpeRiment) is a proposed very large LAr TPC with a well defined 
conceptual design~\cite{Rubbia:2004tz,Rubbia:2009md}, and some of its characteristics are outlined in the rest of 
the paper.  
The underground localization of the experiment is being investigated along the JPARC neutrino beam in 
Collaboration with KEK (Japan) and in Europe within the LAGUNA design study \cite{LAGUNA}.

\section{The GLACIER detector concept}

The proposed cryostat for GLACIER is a single module cryo-tank based on industrial liquefied natural gas (LNG) 
technology. The shape is cylindrical, which gives an excellent surface-to-volume ratio. The detector design is highly 
scalable, up to a total mass of 100 kton. One key technical challenge is to drift free electrons in liquid argon over a 
length as large as  20 meters, which requires special care to achieve and maintain  high purity in the liquid argon
bulk and for the high voltage necessary to generate the drift field (in the order of a MVolt). 
The readout of the charge relies on the novel Large Electron Multiplier (LEM) technique: LAr LEM-TPCs operate in 
double phase with charge extraction and amplification in the vapor phase \cite{Badertscher:2009av}. 
The concept has been very successfully demonstrated on small prototypes: ionization electrons, after drifting in 
the LAr volume, are extracted by a set of grids into the gas phase and driven into the holes of one or more LEMs, 
where charge amplification occurs. 
Each LEM is a thick macroscopic hole multiplier, which can be manufactured with standard PCB techniques. The 
electrons signal is readout via two orthogonal coordinates. One possible configuration, which has been tested
successfully, is to use the induced signal on the segmented upper electrode of the LEM itself and the other by 
collecting the electrons on a segmented anode. 
The images obtained with the LAr LEM-TPC are of very high quality, as in bubble chambers, owing to the 
charge amplification in the LEM, and have good measured dE/ dx resolution. Compared to LAr TPCs with wires 
immersed in liquid argon, which are hard to scale to very large sizes due to mechanical and capacitance (electronic
noise) issues of the long thin wires and by signal attenuation along the drift direction, the LAr LEM-TPC is an elegant 
solution for very large LAr TPCs with long drift paths and mm-sized readout pitch segmentation.
In GLACIER there is also the possibility to detect the Cherenkov light produced by relativistic particles in liquid argon, 
and allows conceiving to have high Tc superconducting solenoid in liquid argon, in order to obtain a magnetized detector. 
The requirement for the excavation of the underground cavern are particularly favorable when compared to other 
detector concepts, requiring a cavern of about 250'000 m$^3$ and relatively shallow depth (600 m.w.e. overburden).
 
\section{R\&D staged approach}

Effective extrapolation to the 100 kton scale requires concrete and well coordinated R\&D steps~\cite{marchionniyp}.
A ton-scale LAr LEM-TPC detector is being operated at CERN within the CERN RE18 experiment (ArDM, focused
on direct Dark Matter searches \cite{ArDM}). 
In order to prove the performance for neutrino physics, additional dedicated test beam campaigns are being 
considered for a detector with a mass of about 10 tons \cite{6m3}, to test and optimize the readout methods and to assess
the calorimetric performance of such detectors.
Beyond these efforts, a 1 kton-scale device is the appropriate choice for a full engineering prototype of a 100 kton 
detector. The choice of the size for the prototype is the result of two a priori contradictory constraints: (1) having 
the largest possible detector as to minimize the extrapolation to 100 kton (2) having the smallest possible detector
to minimize the cost and time requirement to build it. 
A 1 kton detector, which is a full prototype of the GLACIER design, can be built with a diameter of 12 m and a vertical
drift of 10 m \cite{Rubbia:2009md}. From the point of view of the drift path, a mere factor 2 will be needed to 
extrapolate from the prototype to the 100 kton device. Hence, the prototype will be the practical demonstration
for the long drift of free electrons in realistic conditions. 
At the same time, the rest of the volume scaling from the 1 kon to the 100 kton is achieved by increasing the diameter
to about 70 m, and is relatively straightforward noting that (a) large LNG tanks with similar diameters and aspect
ratios already exist (b) the LAr LEM-TPC readout above the liquid will be scaled from an area of 80 m$^2$ (1 kton)
to 3800 m$^2$ (100 kton). The LAr LEM-TPC is fully modular, and this will not require a fundamental extrapolation of
the principle, but rather only poses technical challenges of production, which can be solved in collaboration with industry.

\begin{figure}[ht]
\begin{center}
\includegraphics[width=10cm]{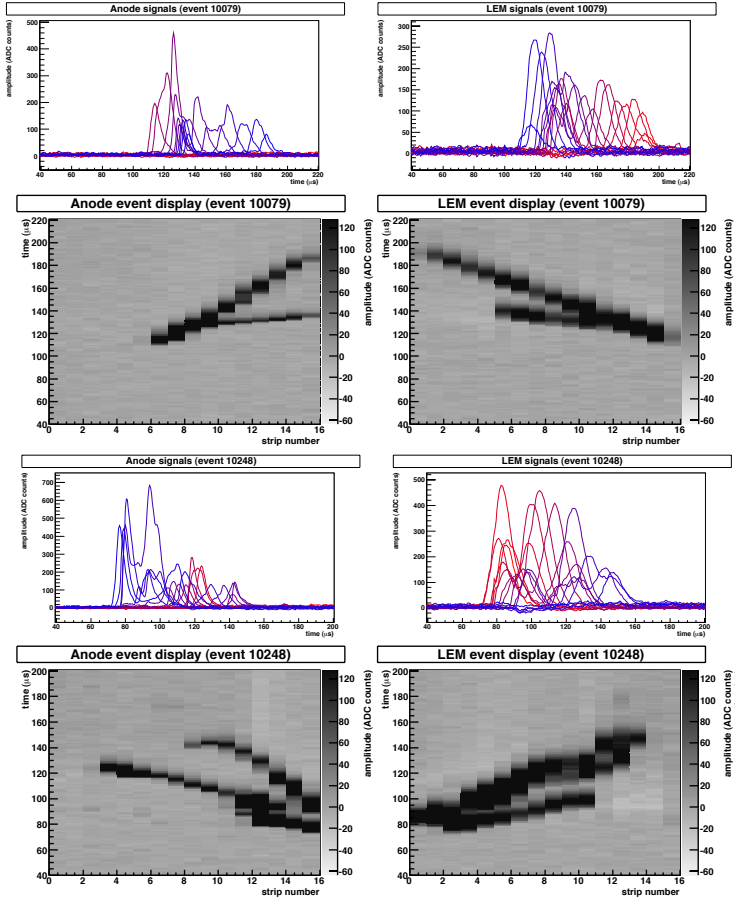}
\caption{Multi-tracks cosmic ray events as seen in the 3 lt LAr LEM-TPC at CERN.}
\label{fig:1}
\end{center}
\end{figure}

\subsection{LEM readout R\&D}

In order to study the properties of the LEM and the possibility to reach high gains in double phase, extensive 
R\&D has been performed on several prototypes, built using standard PCB techniques from different 
manufacturers. Double-sided copper-clad FR4 plates with thicknesses ranging from 0.8 mm to 1.6 mm are drilled
with a regular pattern of 500 $\mu$m diameter holes at a relative distance of 800 $\mu$m.
By applying a potential difference on the two faces of the PCB an intense electric field inside the holes is produced.
A first single stage prototype demonstrated a stable operation in pure Ar at room temperature and pressure up to 
3.5 bar with a gain of 800 per electron. Measurements were performed at high pressure because the density of Ar 
at 3.5 bar is roughly equivalent to the expected density of the vapor at the temperature of 87 K (LAr temperature at 
1 bar). 
Double stage LEM configurations were then tested in pure Ar at room temperature, cryogenic temperature and in 
double phase conditions. The double-stage LEM system demonstrated a gain of $\sim$10$^3$ at a temperature of 
87 K and a pressure of 1 bar. The double-phase operation of the LEM proved the extraction of the charge from the 
liquid to the gas phase.
A 3 lt active volume LAr LEM-TPC was then constructed and successfully operated at CERN. In this setup
a LAr drift volume of 10$\times$10 cm$^2$ cross section and with an adjustable depth of up to 30 cm is followed 
on top by a LEM amplification stage in the Ar vapor at a distance of $\sim$1.5 from the liquid surface. 
Ionization electrons are drifted upward by a uniform electric field generated by a system of field shapers,
extracted from the liquid by means of two extraction grids positioned across the liquid-vapour interface and 
driven to the LEM planes.
The amplified charge is readout by two orthogonal electrodes, one using the induced signal on the segmented upper
electrode of the LEM itself and the other by collecting the electrons on a segmented anode. In this first production both 
readout planes were segmented with 6 mm wide strips, for a total of 32 readout channels for a 10$\times$10 cm$^2$ 
active area. 
A gain of about 10 was achieved in realistic double-phase conditions; two events are shown are shown in 
Fig.~\ref{fig:1}.   
Transverse segmentations down to 2-3 mm will be tested in the near future, together with new configuration 
of the readout electrodes.
As a next step of the LEM readout R\&D, we foresee the development of the full LEM-based readout for a 6m$^3$ LAr 
LEM-TPC designed for test beam exposure.

\section*{Acknowledgements}
We thank the the Swiss Science National Foundation and ETH Z\"urich for
supporting this experimental programme.

\end{document}